\documentclass[twoside,leqno,twocolumn]{article}

\usepackage[letterpaper]{geometry}

\usepackage{ltexpprt}
\usepackage{hyperref}
\usepackage{graphicx}
\usepackage{amsmath}
\usepackage{algorithm}
\usepackage[noend]{algpseudocode}
\usepackage{mathtools}
\usepackage{booktabs}
\usepackage{caption}
\begin{document}

\newcommand\relatedversion{}
\renewcommand\relatedversion{\thanks{The full version of the paper can be accessed at \protect\url{https://arxiv.org/abs/1902.09310}}} 

\title{\Large An Improved A* Search Algorithm for Road
Networks Using New Heuristic Estimation}
\author{Kevin Y. Chen\footnote{Academy for Mathematics, Science, and Engineering, Rockaway, NJ 07866, USA}}

\date{}

\maketitle







\begin{abstract} \small\baselineskip=9pt Finding the shortest path between two points in a graph is a fundamental problem that has been well-studied over the past several decades. Shortest path algorithms are commonly applied to modern navigation systems, so our study aims to improve the efficiency of an existing algorithm on large-scale Euclidean networks. The current literature lacks a deep understanding of certain algorithms' performance on these types of networks. Therefore, we incorporate a new heuristic function, called the $k$-step look-ahead, into the A* search algorithm and conduct a computational experiment to evaluate and compare the results on road networks of varying sizes. Our main findings are that this new heuristic yields a significant improvement in runtime, particularly for larger networks when compared to standard A*, as well as that a higher value of $k$ is needed to achieve optimal efficiency as network size increases. Future research can build upon this work by implementing a program that automatically chooses an optimal $k$ value given an input network. The results of this study can be applied to GPS routing technologies or other navigation devices to speed up the time needed to find the shortest path from an origin to a destination, an essential objective in daily life.\end{abstract}

\section{Introduction}Finding the shortest path from an origin to a destination is a fundamental, well-studied problem faced by researchers and mathematicians for at least the past century. Even in early societies, finding shortest paths was a common, important objective. However, extensive research in the shortest path problem began relatively late as compared to other combinatorial optimization problems, likely due to its apparent simplicity \cite{schrijver_history_2012}. In recent years, these algorithms have been applied to various modern navigation systems such as GPS routing technologies on road networks. 

The problem is typically represented by a graph consisting of nodes and edges, where the goal is to find the shortest path between two given nodes such that the sum of the weights of the traversed edges is minimized. When applying this theoretical model to road networks, nodes often represent intersections, and edges correspond to road segments that are weighted by the length of the segment. Perhaps the most well-documented shortest path algorithm was presented by Dutch computer scientist Edsger W. Dijkstra in 1959 \cite{dijkstra_note_1959}, which remains one of the most efficient methods to this day \cite{zeng_finding_2009}. Dijkstra's algorithm finds the optimal solution to the shortest path problem with nonnegative edge weights by expanding neighboring nodes uniformly in all directions. Another efficient but less-studied technique is the A* (pronounced A-star) algorithm \cite{hart_formal_1968}, which builds off of Dijkstra's by incorporating a heuristic function that takes into consideration the estimated remaining distance from the current node to the destination, thus limiting the search area. 

The heuristic function of the A* search algorithm is a critical part of its implementation because it highly influences the algorithm's runtime and solution optimality \cite{fu_heuristic_2006}. While shortest path problems in general are a well-studied topic, the current literature notably lacks a detailed analysis on how the heuristic estimating method of A* affects its computational performance on real-life networks of various sizes. 

Therefore, the primary objective of this work is to introduce a new heuristic for A* that has not before been implemented and analyzed in the environment of real road networks. This new \textit{look-ahead heuristic} involves the idea of deciding which node in the graph to expand next based on the likelihood that one of the node's children is on the shortest path. We will then evaluate the computational efficiency of this new version of A* and compare it to the standard A* and Dijkstra's algorithm by testing them on road networks of three different sizes. The results of this experiment can be used in the future to determine which of these algorithms is the most optimal to use on Euclidean networks under different circumstances.

\section{Literature Review}While a plethora of past studies have dealt with efforts to compare shortest path algorithms, many of them evaluate their computational performance on randomly generated graphs as opposed to real networks \cite{cherkassky_shortest_1996, biswas_computational_2005, magzhan2013review}. This is problematic when it comes to real-life applications because synthetic graphs lack geographical reality and the most efficient shortest path algorithms on random networks may not practically be ideal when used on empirical systems \cite{zeng_finding_2009}. Furthermore, these studies exclude the A* algorithm in their comparisons, but this decision is somewhat reasonable because A* functions best in the context of Euclidean space instead of randomized networks. While Zhan and Noon \cite{zhan_shortest_1998} examine the performance of multiple shortest path algorithms on several large-scale U.S. road networks, they do not mention A*. Additionally, Chan et al. \cite{chan_experiment_2016} compare the efficiency of six different algorithms including A*, but do so on a relatively small bus route that is not representative of a typical large transportation network.

On the other hand, other previous experiments carry out direct comparisons between A* and Dijkstra's algorithm on complex road networks, but do so without varying certain features of the algorithms \cite{rachmawati_analysis_2020, jain_shortest_2021}. They generally find the standard A* to discover the shortest path faster in those environments. Zeng and Church \cite{zeng_finding_2009} analyze the effects of varying the type of data structure used to maintain the set of candidate nodes (e.g. $k$-array heaps, double buckets, approximate buckets) in A* and Dijkstra's. Moreover, in the previously mentioned papers that study the application of shortest path algorithms to road networks, Euclidean distance is the most commonly used heuristic function in A*, as it ideally describes the lower bound of the true cost from a node to the destination in Euclidean space. Some more studies attempt to incorporate an ``overdo" heuristic into A* that sacrifices solution optimality for a shorter runtime \cite{engineer_fast_2001, jacob_computational_1999}. Finally, others explore a variation of A* known as the ALT algorithm, which uses a different lower-bounding technique based on landmarks and the triangle inequality as opposed to Euclidean bounds \cite{goldberg_computing_2005, goldberg_reach_2006}.

To fill in the aforementioned gaps in the literature, our main contribution is to improve A*'s performance on large road networks by integrating a heuristic function that has the potential to decrease runtime while guaranteeing the optimal path.

\section{Purpose}The goal of this study is to improve upon the performance of the A* search algorithm on Euclidean road networks by implementing a $k$-step look-ahead heuristic. Specifically, we break this problem down into the following steps:

\begin{enumerate}
    \item We aim to evaluate the computational performance of Dijkstra's algorithm, the standard A* algorithm, and the A* algorithm with a $k$-step look-ahead heuristic function on three real road networks of different sizes. We compare the efficiency of these algorithms in terms of program runtime.
    \item We will then use the results of the computational experiment to determine which of the algorithms is best to use for different sizes of the shortest path problem on Euclidean networks.
\end{enumerate}


\section{Shortest Path Algorithms}

In the past, scientists and mathematicians have derived various algorithms for solving the shortest path problem. In this paper, we focus on evaluating Dijkstra's algorithm and the A* algorithm, and we provide an overview of each of these methods in the following sections.

\subsection{Notation and Definitions}

In this section, we discuss the notation necessary for understanding graphs and shortest path algorithms. Most of the notation introduced here will be used throughout section 4. We then rigorously define the shortest path problem.

Let $G = (V, E)$ be a \textit{directed, undirected,} or \textit{mixed graph} consisting of a set of \textit{vertices} (i.e. \textit{nodes}) $V$ and a set of \textit{edges} $E$. We denote $e_{x,y} \in E$ as the edge from node $x$ to node $y$, and $c(x, y)$ as the \textit{cost} (i.e. \textit{weight}) of that edge. We also say that a node $n'$ is \textit{adjacent} to node $n$ if there exists an edge $e_{n,n'}$. In the context of shortest path algorithms, we let $d(n)$ be the total cost of the shortest path from the source node to node $n$. We also call this value the \textit{distance value} of node $n$. We let the \textit{source node} of the shortest path problem be $s$ and the \textit{target node} be $t$. We also have $\textsf{dist}(x,y)$ represent the Euclidean (i.e. straight-line) distance between nodes $x$ and $y$. 

Next, let \[P = ((n_1,n_2),(n_2,n_3),\dots,(n_{m-1},n_m))\] be a \textit{path} of edges connecting node $n_1$ to $n_m$ where each $(n_i,n_{i+1})$ is an edge of the path connecting adjacent nodes $n_i$ and $n_{i+1}$ for all $1 \le i < m$. We say that a path $P$ is \textit{valid} if and only if $y_j = x_{j+1}$ for all $1 \le j < m-1$, where $(x_j,y_j)$ and $(x_{j+1},y_{j+1})$ are the $j$th and $(j+1)$th elements of $P$, respectively. We define a path to be \textit{acyclic} if the path does not pass through any same node more than once. We call node $n'$ a \textit{successor} of $n$ if $n'$ comes after $n$ in a given path. In addition, a path between node $n$ and $n'$ is said to have \textit{length} $k$ if the path passes through exactly $k$ edges (note that this is a different quantity from total cost of the path).

The goal of the \textit{single-pair shortest path problem} is to find the path $P$ between $s$ and $t$, where $n_1=s$ and $n_m=t$, that minimizes the sum $\sum_{i=1}^{m-1} {c(n_i,n_{i+1})}$ over all possible values of $m$. Other common variations of the shortest path problem exist besides the single-pair problem. These include the single-source problem, where the objective is to find the shortest paths from one node to all other nodes, and the all-pairs problem, where a shortest path must be found between all pairs of nodes in the graph \cite{madkour2017survey}.

\subsection{Dijkstra's Algorithm}

First introduced by Dijkstra in 1959 \cite{dijkstra_note_1959}, this algorithm finds the optimal path on a graph with nonnegative weights from a single source node to a target or set of targets. While the algorithm is more commonly used to solve the single-source shortest path problem, it will be implemented in this study to handle a single pair of nodes for the purpose of comparing its computational efficiency with that of A*.

Dijkstra's method, as outlined in Algorithm 1, utilizes a min-priority queue to store the set of unvisited nodes, leading to more efficient computations compared to other basic data structures \cite{zeng_finding_2009}. Since our work focuses on finding the shortest path from an origin to a particular destination, the algorithm terminates when the target node $t$ is marked as visited. Asymptotically, the algorithm has a worst-case time complexity of $O((|V| + |E|) \log |V|)$ when a standard binary heap is implemented as the priority queue data structure \cite{cormen_introduction_2009}.

\begin{algorithm}[H]
\caption{Dijkstra's Algorithm}
\vspace{0.2cm}
\noindent\underline{$\textsf{Dijkstra}(G, s, t)$}:
\vspace{0.1cm}\newline
\textbf{Input:} A graph $G = (V,E)$, start node $s \in V$ and target node $t \in V$.\\
\textbf{Output:} The shortest path of edges from node $s$ to node $t$.
\begin{algorithmic}[1]
\State $\textsf{unvisitedQueue} \coloneqq$ priority queue of unvisited nodes sorted by $d$ value
\State $\textsf{previous} \coloneqq$  an empty map
\State $d(n) \coloneqq \infty$ for all nodes $n$ in $G$
\State $d(s) \coloneqq 0$
\While{\textsf{unvisitedQueue} not empty and $t$ not visited}
    \State Extract node $i$ from $\textsf{unvisitedQueue}$
    \For{all $j$ adjacent to $i$}
        \If{$j$ is visited}
            \State Continue
        \EndIf
        \State $\textsf{dTemp} \coloneqq d(i) + c(i,j)$
        \If{$\textsf{dTemp} < d(j)$}
            \State $d(j) \coloneqq \textsf{dTemp}$
            \State $\textsf{previous}[j] \coloneqq i$
        \EndIf
    \EndFor
    \State Reorder \textsf{unvisitedQueue}
\EndWhile
\vspace{0.2cm}
\end{algorithmic}
\end{algorithm}

\subsection{A* Algorithm}

The A* search algorithm is a best-first search method \cite{russell_artificial_2010} and has a general procedure similar to Dijkstra's, but the difference comes with the concept of a heuristic function that estimates the remaining distance from node $n$ to the target node $t$. This crucial feature causes A* to run faster than Dijkstra's in many past computational experiments \cite{zeng_finding_2009, rachmawati_analysis_2020, engineer_fast_2001}. A* functions by choosing which path to extend based on the node that minimizes the value of \[f(n) = g(n) + h(n),\] where $n$ is the next node in the path, $g(n)$ is the current distance value of $n$, and $h(n)$ is the predetermined heuristic value of $n$. Using this heuristic, A* is able to concentrate its search in the direction of the target by prioritizing nodes with a lower $h$ value (i.e. nodes that are estimated to be closer to the destination). This main distinction between A* and Dijkstra's can be seen in Algorithm 2.

\begin{algorithm}[H]
\caption{A* Search Algorithm}
\vspace{0.2cm}
\noindent\underline{$\textsf{A*}(G, s, t, h)$}:
\vspace{0.1cm}\newline
\textbf{Input:} A graph $G = (V,E)$, start node $s \in V$, target node $t \in V$, and heuristic function $h(n)$. \\
\textbf{Output:} The shortest path of edges from node $s$ to node $t$.
\begin{algorithmic}[1]
\State $\textsf{unvisitedQueue} \coloneqq$ priority queue of unvisited nodes sorted by $f$ value
\State $\textsf{previous} \coloneqq$  an empty map
\State $f(n) \coloneqq \infty$ for all nodes $n$ in $G$
\State $g(n) \coloneqq \infty$ for all nodes $n$ in $G$
\State $f(s) \coloneqq h(s)$
\State $g(s) \coloneqq 0$
\While{\textsf{unvisitedQueue} not empty and $t$ not visited}
    \State Extract node $i$ from $\textsf{unvisitedQueue}$
    \For{all $j$ adjacent to $i$}
        \If{$j$ is visited}
            \State Continue
        \EndIf
        \State $\textsf{gTemp} \coloneqq g(i) + c(i,j)$
        \If{$\textsf{gTemp} < g(j)$}
            \State $g(j) \coloneqq \textsf{gTemp}$
            \State $f(j) \coloneqq g(j) + h(j)$
            \State $\textsf{previous}[j] \coloneqq i$
        \EndIf
    \EndFor
    \State Reorder \textsf{unvisitedQueue}
\EndWhile
\vspace{0.2cm}
\end{algorithmic}
\end{algorithm}

The most commonly used heuristic estimator when dealing with real networks is Euclidean distance, as this metric always describes the lowest bound of the distance between any two points in space \cite{sharma_shortest_2015, schultes_route_2008}. Euclidean distance is an example of an \textit{admissible heuristic}, which is defined as a heuristic that never overestimates the true lowest possible cost from a node $n$ to $t$, thereby guaranteeing that the shortest path to $t$ is not overlooked. Additionally, a heuristic is considered \textit{consistent} (or \textit{monotone}) if it satisfies \[h(n) \le c(n,p) + h(p) \text{ and } h(t) = 0,\] where $p$ is a successor of any node $n$ in the graph. In the context of A*, this means that the cost by which a node is reached must be the lowest possible once that node is expanded, given that the graph only includes nonnegative edge weights. When the heuristic is simply Euclidean distance, this property of consistency is analogous to the triangle inequality because it essentially states that the straight-line distance between $n$ and $t$ is never greater than the sum of the distance between $n$ and $p$ plus the distance between $p$ and $t$. A consistent heuristic is always admissible, but the converse is not necessarily true. This can be proven by induction \cite{dechter_generalized_1985}. 

While the time complexity of A* depends heavily on the search space and the heuristic function used, its worst-case performance on an unbounded search space is $O(b^d)$, where $b$ is the branching factor (i.e. the average number of children per node, or the outdegree) and $d$ is the depth of the shortest path solution \cite{russell_artificial_2010}.

\subsection{The $k$-step Look-ahead Heuristic}

We aim to improve the computational efficiency of the A* search algorithm by integrating the $k$-step look-ahead heuristic, in which $k$ is a parameter that we choose. First, we start with a couple of definitions. We define $C(P,t)$ to be a \textit{cut function} that represents the section of path $P$ up until node $t$, if $t$ is on path $P$. For instance, if $P = ((a,b),(b,t),(t,c),(c,d))$, then $C(P,t) = ((a,b),(b,t))$. If $P$ does not pass through $t$, then $C(P,t) = P$. If $X$ is the set of all visited nodes in $G$, then let $S_{G,X,n,k}$ be the set of all acyclic paths from $n$ of length $k$ that do not pass through any nodes in $X$.

The intuitive explanation behind the $k$-step look-ahead heuristic is that the heuristic scans next possible paths to a certain extent from a given node and selects the estimated cost of the path that seems most promising. Specifically, instead of estimating the remaining cost to get from node $n$ to target node $t$ directly using Euclidean distance, we do so by iteratively finding the cost of each possible path of length $k$ to the unvisited successors of $n$, added to the Euclidean distance between that successor node and $t$. We then select the lowest of those calculated values and assign that to $h(n)$. More formally, we have \[h(n) = \min_{P \in S_{G,X,n,k}} \left[ \left( \sum\limits_{(x,y)\in C(P,t)}c(x,y) \right) + \textsf{dist}(n',t) \right]\] where $n'$ is the final node in path $C(P,t)$. 

For example, to find the estimated cost from $n$ to $t$ using the 1-step look-ahead heuristic, we calculate \[c(n,n') + \textsf{dist}(n',t)\] for every unvisited child node $n'$ of $n$, and assign $h(n)$ to the smallest resulting value.

We emphasize two details implied from our definition of the $k$-step look-ahead heuristic that become increasingly important as the value of $k$ gets larger:

\begin{enumerate}
    \item As previously mentioned, the ``look-ahead" path $P$ that our algorithm takes must be acyclic, i.e. it should not revisit any nodes previously passed through earlier in the same look-ahead path. Otherwise, the process could be simulating a sub-optimal path where nodes are revisited, potentially leading to unnecessary calculations. This is particularly true when dealing with undirected graphs where it is possible to travel between nodes in both directions. The list \textsf{alreadySeen} in Algorithm 3 keeps track of which nodes have already been looked through in a given look-ahead path.
    \item If the algorithm reaches the target node during its look-ahead process, that particular look-ahead path should terminate and the returned value should be the cost of that path from node $n$ to $t$. This prevents the algorithm from unnecessarily extending to successor nodes beyond the desired target node. This is the reasoning behind finding the cost of path $C(P,t)$ as opposed to the entire cost of $P$.
\end{enumerate}

Algorithm 3 shows example pseudocode for computing the 2-step look-ahead heuristic value of a node $n$.

\begin{algorithm}[H]
\caption{2-step Look-Ahead Heuristic}
\vspace{0.2cm}
\noindent\underline{$\textsf{2-step Look-ahead}(G, n, t)$}:
\vspace{0.1cm}\newline
\textbf{Input:} A Euclidean graph $G = (V,E)$, node $n \in V$, and target node $t \in V$. \\
\textbf{Output:} $h(n)$, or the estimated cost of the shortest path from $n$ to $t$.
\begin{algorithmic}[1]
\If {$n = t$}
    \State $h(n) \coloneqq 0$
\Else
    \State $\textsf{alreadySeen} \coloneqq$ an empty list only containing $n$
    \State $\textsf{possibleVals} \coloneqq$ an empty list only containing $\infty$
    \For {all $n_1$ adjacent to $n$}
        \If {$n_1$ has not been visited}
            \If{$n_1 = t$}
                \State Add $c(n,n_1)$ to \textsf{possibleVals}
                \State Continue
            \EndIf
            \State Add $n_1$ to \textsf{alreadySeen}
            \For{all $n_2$ adjacent to $n_1$}
                \If{$n_2$ is not visited and not in \textsf{alreadySeen}}
                    \State Add $c(n,n_1) + c(n_1,n_2) + \textsf{dist}(n_2,t)$ to \textsf{possibleVals}
                \EndIf
            \EndFor
            \State Remove $n_1$ from $\textsf{alreadySeen}$
        \EndIf
    \EndFor
    \State $h(n) \coloneqq \min$(\textsf{possibleVals})
\EndIf
\State \Return $h(n)$
\vspace{0.2cm}
\end{algorithmic}
\end{algorithm}

The $k$-step look-ahead is a more accurate heuristic of the shortest path between $n$ and $t$ than direct Euclidean distance because it takes into account some of the edges that the real shortest path from $s$ to $t$ will actually traverse. This suggests that a shortest path will be found quicker with this new heuristic than by using standard A*, particularly for relatively small values of $k$ (for larger $k$, the number of calculations the heuristic has to make tends to increase exponentially, which can lead to slower runtimes). Furthermore, this estimator is still admissible, as the resulting value for $h(n)$ will never be greater than the cost of the true shortest path from $n$ to $t$, so the optimal solution is guaranteed on Euclidean networks. 

\section{Methodology}

In this section, we discuss the procedure used to perform our computational experiment on the efficiency of Dijkstra's algorithm, standard A*, and A* with a $k$-step look-ahead heuristic for various values of $k$ on sample road networks. To our knowledge, the new heuristic has not before been evaluated on Euclidean networks. All tests are run using the Python programming language \cite{hong_dijkstras_2020} in Google Colaboratory on a computer with a 1.3 GHz Intel Core i7 processor and 16 GB of RAM.

\subsection{Test Network Details}

We implement the shortest path algorithms on three different road networks as described in Table \ref{tab:networks}: the relatively small Downtown Brooklyn network, the medium-sized Jersey City network, and the larger New York City network (see Appendix A for visual layouts). We choose to include datasets of different sizes for the purpose of determining which algorithms perform comparatively well or poorly for various scales of networks. The data is obtained from the OSMnx Street Networks Dataverse, a compilation found on Harvard Dataverse that includes the street networks of every city, town, urban area, county, census tract, and Zillow-defined neighborhood in the United States \cite{DVN/CUWWYJ_2017, boeing2020multi}. The networks are created from OpenStreetMap data using the OSMnx software \cite{boeing2017osmnx}. In the networks, nodes correspond to street intersections, edges represent the road segments connecting those intersections, and weights correspond to the lengths of those road segments in meters. Additionally, for each network, the data source provides node and edge lists. The node lists contain each node in the network, its assigned ID, and its geographic coordinates, while the edge lists include information about the two nodes that each edge connects, the cost of the edge, and whether the edge is one-way only, among other features. Note that these networks are considered mixed graphs because they contain both directed and undirected edges.

\begin{table}
    \centering
    \caption{Characteristics of test road networks.}
    \begin{tabular}{lll}
    \toprule
    Road network      & \# of nodes & \# of edges \\ \midrule
    Downtown Brooklyn & 149         & 281         \\
    Jersey City       & 2232        & 5320        \\
    New York City     & 55136       & 141085      \\ \bottomrule
    \end{tabular}
    \label{tab:networks}
\end{table}

Each road network dataset must also be pre-processed in order to remove groups of nodes that are entirely disconnected from the rest of the network, thus ensuring a fully navigable road system. Lastly, before the final experiment is run, all algorithms are tested on several source-target pairs for all three networks to confirm that they consistently find the same optimal path.

\subsection{Algorithm Implementations}

In our computational experiment, twelve algorithms are tested on each network: Dijkstra's, standard A*, and A* with a $k$-step look-ahead heuristic where $k$ is each of 1 to 10 inclusive. For each network, a certain number of source-target node pairs are randomly selected from the network (with the exception of New York City), and each algorithm is run a fixed number of times for each source-target pair. Specifics are outlined in Table \ref{tab:experiment}. We then record the resulting average runtime and average nodes expanded (i.e. nodes visited) for an algorithm's execution (see Results section). As Table \ref{tab:experiment} shows, we design our experiment differently between road networks in terms of the number of times each algorithm is run. This is because the overall runtimes of the algorithms significantly increase with the size of the network, and runtimes on the smaller networks must be recorded at a greater degree of precision. Since we run the algorithms on only one distinct shortest path problem in the New York City dataset, we deliberately choose the source and destination nodes to be on opposite sides of the network in order to encapsulate the network's size and complexity.

\begin{table}
    \centering
    \caption{Experimental details for each test road network.}
    \resizebox{8cm}{!}{
    \begin{tabular}{lll}
    \toprule
    Road Network      & \# of source-target pairs & \# of runs per pair \\ \midrule
    Downtown Brooklyn & 200         & 30     \\
    Jersey City       & 100         & 10     \\
    New York City     & 1           & 1      \\ \bottomrule
    \end{tabular}
    \label{tab:experiment}
}
\end{table}

To implement a priority queue data structure to efficiently store unvisited nodes, we utilize Python's \textit{heapq} module, which represents a binary tree and can order a list by priority in linear time. Moreover, we incorporate a calculation of Euclidean distance between two given nodes in meters using the haversine formula since the geographic coordinates of each node is known \cite{jain_shortest_2021}. The formula states that

\resizebox{8cm}{!}{$d = 2r \arcsin \left( \sqrt{\sin^2 \left( \dfrac{\phi_2 - \phi_1}{2} \right) + \cos(\phi_1)\cos(\phi_2) \sin^2 \left( \dfrac{\lambda_2 - \lambda_1}{2} \right) } \right),$}

where 

\begin{itemize}
    \item $(\phi_1, \lambda_1)$ and $(\phi_2, \lambda_2)$ are the coordinates of the two points in radians,
    \item $r$ is the radius of the Earth, and
    \item $d$ is the great-circle distance between the two points.
\end{itemize}

Finally, for our implementations of all variations of A*, the final recorded runtimes do not include the program's graph pre-processing steps of calculating the Euclidean distance between each node in the network and the target node.

\section{Results}

The results of our computational experiment with shortest path algorithms on road networks are displayed in this section. As previously described, we run the A* algorithm with a $k$-step look-ahead heuristic on each of the three sample road networks for all $k$ from 1 to 10 inclusive. Additionally, we include Dijkstra's algorithm to provide context of the algorithms' performance. As is a common convention \cite{zeng_finding_2009, zhan_shortest_1998, chan_experiment_2016, rachmawati_analysis_2020}, we use runtime to judge an algorithm's efficiency. For each source-target pair, all algorithms we implement guarantee the same shortest path. See Appendix B for additional visualizations of the number of nodes expanded as a function of $k$.

\subsection{Small Sample Network}

\begin{figure}
    \centering
    \includegraphics[width=8cm]{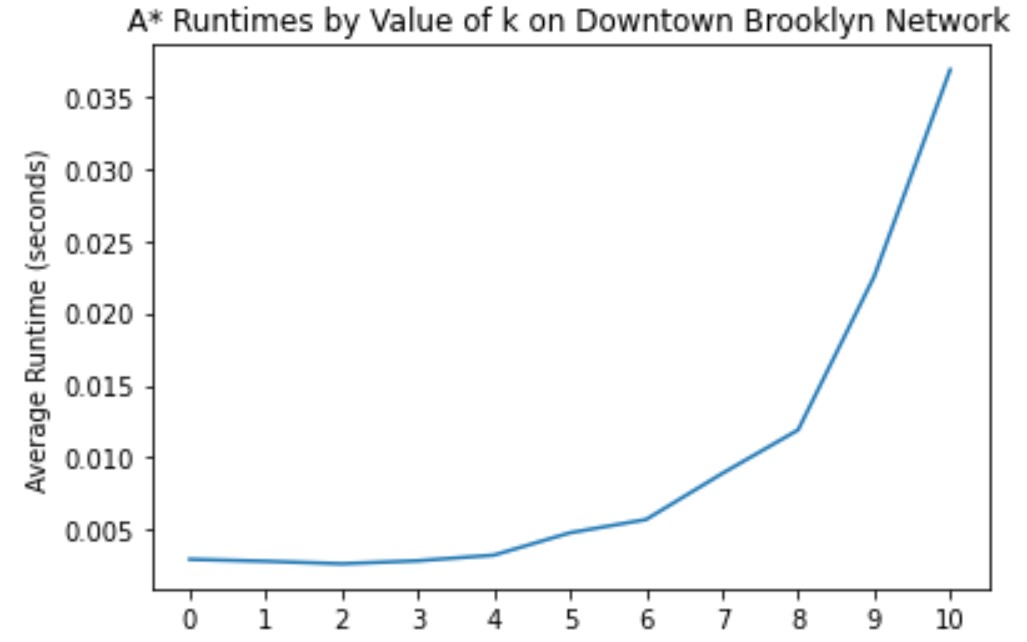}
    \caption{A visualization of the average algorithm runtime in seconds as a function of $k$ in A* implementation with a $k$-step look-ahead heuristic when run on the Downtown Brooklyn network. A value of $k = 0$ indicates the standard A* algorithm. Optimal efficiency for this network is achieved with a 2-step look-ahead, but its improvement from standard A* is negligible.}
    \label{fig:dtb}
\end{figure}

Results for all twelve algorithms on the Downtown Brooklyn street network are shown in Table \ref{tab:dtb}. In terms of runtime, a 2-step look-ahead heuristic achieves the greatest efficiency by a small margin for this particular network at roughly 0.0026 seconds on average and 10.5\% faster than standard A*. Moreover, the heuristics with $k=1$ and $k=3$ also perform better than regular A*. Figure \ref{fig:dtb} shows that as $k$ increases, A*'s average runtime grows exponentially. While $k=2$ achieves the quickest runtime, A* expands the least number of nodes on average with a 4-step look-ahead. In addition, when $k=10$, the largest value of $k$ that we test, the runtime is significantly greater than that of Dijkstra's while still expanding less than half as many nodes.

\begin{table}
\centering
\caption{Comparison of Algorithm Performance on Downtown Brooklyn Network.}
\resizebox{8cm}{!}{
\begin{tabular}{lll}
\toprule
Algorithm   & Avg Runtime (s)   & Avg \# of Nodes Expanded \\ \midrule
Dijkstra's  & 0.0055            & 69.81    \\
Standard A* & 0.0030            & 34.44    \\
A*, $k=1$   & 0.0028            & 30.59    \\
A*, $k=2$   & 0.0026            & 27.83    \\
A*, $k=3$   & 0.0029            & 26.92    \\
A*, $k=4$   & 0.0032            & 25.59    \\
A*, $k=5$   & 0.0048            & 26.91    \\
A*, $k=6$   & 0.0057            & 28.98    \\
A*, $k=7$   & 0.0089            & 29.39    \\
A*, $k=8$   & 0.0119            & 29.34    \\
A*, $k=9$   & 0.0225            & 31.28    \\
A*, $k=10$  & 0.0368            & 30.68    \\ \bottomrule
\end{tabular}
}
\label{tab:dtb}
\end{table}

\subsection{Medium Sample Network}

\begin{figure}
    \centering
    \includegraphics[width=8cm]{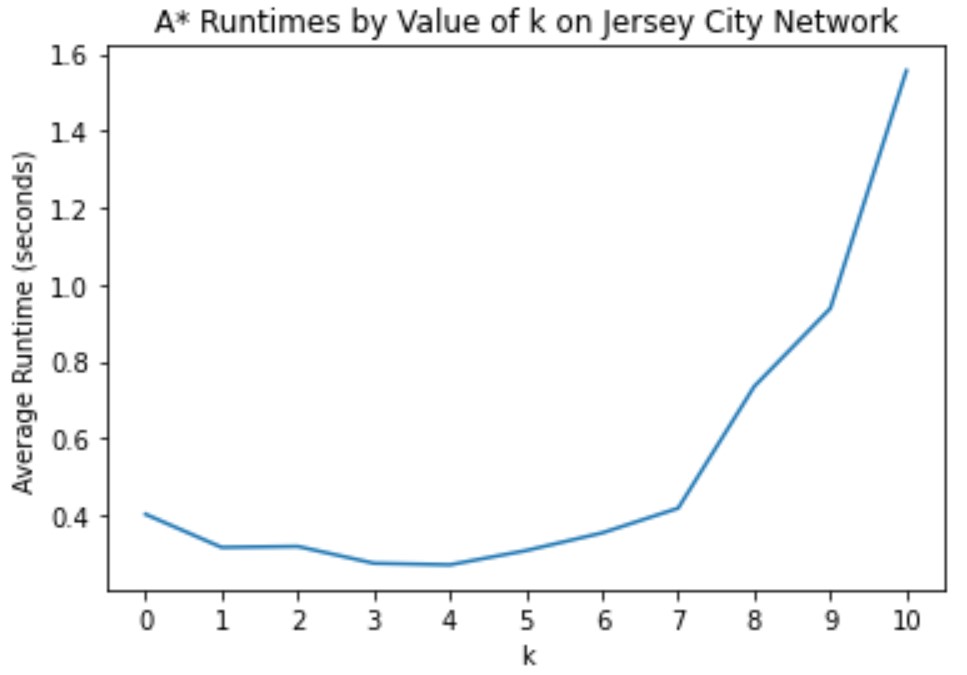}
    \caption{Algorithm runtime in seconds as a function of $k$ in our implementation of A* search algorithm with a $k$-step look-ahead heuristic when run on the Jersey City network. A* achieves optimal efficiency using a 4-step look-ahead heuristic for this particular network.}
    \label{fig:jc}
\end{figure}

As shown in Table \ref{tab:jc}, the A* algorithm efficiency is able to be improved for 6 different values of $k$ on the medium-sized Jersey City road network. Most notably, the best performance occurs when $k=4$, running 32.6\% faster than regular A* at approximately 0.2719 seconds on average. This same value of $k$ also results in the least average number of nodes expanded. We also see a growing exponential behavior in algorithm runtime for values of $k$ greater than $4$, depicted in Figure \ref{fig:jc}. Once $k$ hits 10, A*'s runtime exceeds that of Dijkstra's even though it expands much fewer nodes.

\begin{table}
\centering
\caption{Comparison of Algorithm Performance on Jersey City Network.}
\resizebox{8cm}{!}{
\begin{tabular}{lll}
\toprule
Algorithm   & Avg Runtime (s)   & Avg \# of Nodes Expanded \\ \midrule
Dijkstra's  & 1.1554            & 1192.50    \\
Standard A* & 0.4033            & 304.88    \\
A*, $k=1$   & 0.3172            & 247.29    \\
A*, $k=2$   & 0.3196            & 220.90    \\
A*, $k=3$   & 0.2759            & 203.58    \\
A*, $k=4$   & 0.2719            & 192.81    \\
A*, $k=5$   & 0.3088            & 210.99    \\
A*, $k=6$   & 0.3547            & 220.63    \\
A*, $k=7$   & 0.4190            & 233.78    \\
A*, $k=8$   & 0.7356            & 240.70    \\
A*, $k=9$   & 0.9383            & 254.02    \\
A*, $k=10$  & 1.5565            & 261.78    \\ \bottomrule
\end{tabular}
\label{tab:jc}
}
\end{table}

\subsection{Large Sample Network}

\begin{figure}
    \centering
    \includegraphics[width=8cm]{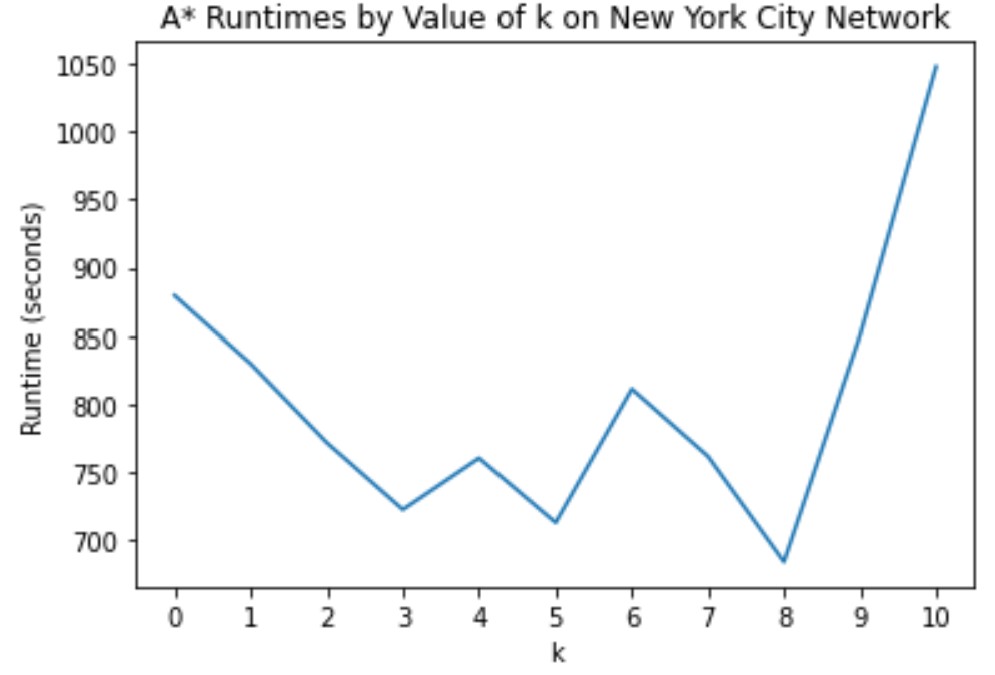}
    \caption{Algorithm runtime in seconds as a function of $k$ for A* with a $k$-step look-ahead heuristic when run on the New York City network. A* achieves optimal efficiency using an 8-step look-ahead heuristic for this particular network, but also performs relatively well when $k=3$ and $k=5$.}
    \label{fig:nyc}
\end{figure}

The street system of New York City is by far the largest of the three sample networks we use in this experiment, and all runtimes surpass 10 minutes. We test each algorithm once on a chosen source-target pair in this network, and find that an 8-step look-ahead heuristic achieves optimal performance at just 684.10 seconds, as indicated in Table \ref{tab:nyc}. Using this heuristic results in a 22.2\% shorter runtime than using standard A*. In total, 9 different values of $k$ perform better than regular A*. Figure \ref{fig:nyc} visualizes this data, showing that the runtimes stay relatively low when $k$ is between 3 and 8, but skyrockets once $k$ exceeds 8. Additionally, A* expands the least number of nodes in the New York City network when using a 4-step look-ahead heuristic. When $k=10$, A* runs more efficiently than Dijkstra's while also expanding less than half as many nodes.

\begin{table}
\centering
\caption{Comparison of Algorithm Performance on New York City Network.}
\resizebox{8cm}{!}{
\begin{tabular}{lll}
\toprule
Algorithm   & Runtime (s)       & \# of Nodes Expanded \\ \midrule
Dijkstra's  & 1389.98           & 54346    \\
Standard A* & 879.43            & 23370    \\
A*, $k=1$   & 829.41            & 20498    \\
A*, $k=2$   & 771.66            & 19200    \\
A*, $k=3$   & 722.54            & 18323    \\
A*, $k=4$   & 760.31            & 17769    \\
A*, $k=5$   & 712.97            & 17934    \\
A*, $k=6$   & 811.00            & 18959    \\
A*, $k=7$   & 761.65            & 20418    \\
A*, $k=8$   & 684.10            & 20559    \\
A*, $k=9$   & 850.23            & 20264    \\
A*, $k=10$  & 1047.57           & 20348    \\ \bottomrule
\end{tabular}
}
\label{tab:nyc}
\end{table}

\section{Discussion}

Through computational experimentation, our goal has been to determine whether implementing a $k$-step look-ahead heuristic would improve the efficiency of the A* search algorithm and to decide which algorithms work best on different sizes of road network. Here, we address this problem by interpreting the results from the previous section.

First, the results agree with our hypothesis that the $k$-step look-ahead heuristic would lead to shorter runtime than using direct Euclidean distance as the estimator. For the Downtown Brooklyn network, although we are able to slightly improve efficiency for a few values of $k$, the differences in runtime are negligible ten-thousandths of a second, which is unlikely to be impactful in the real world. On the other hand, using the new heuristic on the Jersey City and New York City networks shortens runtime by at most 0.13 and 195.33 seconds, respectively. In addition, on the Downtown Brooklyn network, 3 values of $k$ improve the efficiency of A*, whereas this number is 6 for the Jersey City network and 9 for New York City. The optimal efficiency occurs when $k=2$ for Downtown Brooklyn, $k=4$ for Jersey City, and $k=8$ for New York City. This suggests not only that the $k$-step look-ahead heuristic is more effective for larger networks, but that the single optimal value of $k$ increases as the network size grows. The above conclusions seem reasonable because in a larger graph, there is a greater need to expand an algorithm's search in the direction of the target node, which is accomplished by greater values of $k$ to create a more accurate heuristic. 

A reason that one shortest path algorithm may run faster than another may be that it visits and expands less nodes, but this metric does not tell the whole story in all cases. For example, in each of the two smaller networks (i.e. Downtown Brooklyn and Jersey City), A* with a 10-step look-ahead heuristic has a considerably longer runtime than Dijkstra's algorithm, but the number of nodes that A* expands in these cases is significantly lower than that of Dijkstra's. This is explained by the sheer number of steps that A* has to take when calculating the cost of each look-ahead path. As mentioned earlier, the number of steps needed to calculate a $k$-step look-ahead heuristic value grows exponentially as $k$ increases due to the many possible look-ahead paths that diverge at each node.

Meanwhile, the situation is different with the New York City network. In this case, both the runtime and number of nodes expanded are less for A* with a 10-step look-ahead than for Dijkstra's. This further emphasizes A*'s suitability for large-scale networks, even for relatively high values of $k$.

A more surprising result of this experiment emerges when looking at data for the number of nodes expanded. The original speculation is that a greater value of $k$ leads to a more accurate heuristic, thus expanding fewer ``incorrect" nodes that are not on the shortest path. However, this does not seem to be the case. In particular, for all three networks, the minimal number of nodes expanded during A* search occurs when $k=4$. Finding a justification to explain this phenomenon can be one direction of future work. Another common theme between all three test networks is that all variations of A* expand significantly fewer nodes than Dijkstra's does, likely due to Dijkstra's tendency to blindly expand nodes in all directions until the destination is found, as it is not guided by a heuristic. For instance, in the New York City network, Dijkstra's algorithm expands 54346 out of 55136 nodes in the entire graph, while all A* implementations go about finding the shortest path more intelligently. Note that the reason Dijkstra's expands nearly all the nodes in the network for this specific case is that we deliberately choose the source and target nodes to be on opposite ends of the network.

While many useful insights are gained from our interpretations, these are preliminary results that may need more testing to solidify. For example, our single run on the New York City dataset for each algorithm may not be fully representative of all possibilities on the network; nevertheless, our analysis in this section is likely safe to generalize and apply to the real world. To further strengthen the validity of our conclusions, future experiments should run these algorithms on many more source-target pairs on several different road networks.

Overall, the results of this study show that incorporating a $k$-step look-ahead heuristic into the A* search algorithm yields greater efficiency in solving the shortest path problem on transportation networks and can be particularly effective on larger, more complex road systems.

\section{Conclusion and Future Work}

The shortest path problem is a fundamental combinatorial optimization problem that has been intensely studied for the past several decades. In the current literature, few studies have compared various shortest path algorithms' performance on large, complex road networks \cite{zeng_finding_2009, zhan_shortest_1998, engineer_fast_2001}. The main contribution of our work is to implement the A* search algorithm with a $k$-step look-ahead heuristic to solve the single-pair shortest path problem on various road networks for the first time. We implement Dijkstra's algorithm, A*, and A* with a $k$-step look-ahead heuristic for all $k$ from 1 to 10 inclusive. We then compare these algorithms' performance on three real road networks of different sizes: the small Downtown Brooklyn network, the medium-sized Jersey City network, and the larger New York City network. From the results of our experiment, we conclude that using the look-ahead heuristic significantly improves the efficiency of A*, especially on larger networks. We also find that the greater the network size, the higher $k$ needs to be in order to achieve optimal efficiency on the given network. Limitations of our experimental methodology include the lack of a diverse range of source-target pairs to fully represent all possible shortest path problems on the New York City network. The results of this study can be incorporated into cutting-edge navigation systems on transportation networks, such as GPS routing technologies, as finding the shortest possible path between two locations is a common everyday objective.

Some possibilities of related future research directions include incorporating different data structures (i.e. double buckets, approximate buckets) for the priority queue in the implementation of A* with $k$-step look-ahead, as opposed to the simple binary heap structure we use in this study. Additionally, the average degree of a graph (i.e. the average number of edges that are incident to each node in the graph) may be another factor that affects the efficiency of using a $k$-step look-ahead heuristic, and should be analyzed in the context of real road networks. Finally, future studies may explore the possibility of implementing a program that automatically chooses the most optimal value of $k$ to use in A*. The efficiency of this algorithm can then be evaluated on various Euclidean networks.

\section{Acknowledgements}

The author thanks Jesse Stern (Department of Computer Science, University of Chicago) for feedback and mentorship on this project. The author also thanks the Summer STEM Institute (SSI).

\bibliography{mybib}{}
\bibliographystyle{ieeetr} 

\newpage
~\newpage
\appendix
\section{Sample Road Networks}

\begin{figure}[h!]
    \centering
    \includegraphics[width=7cm]{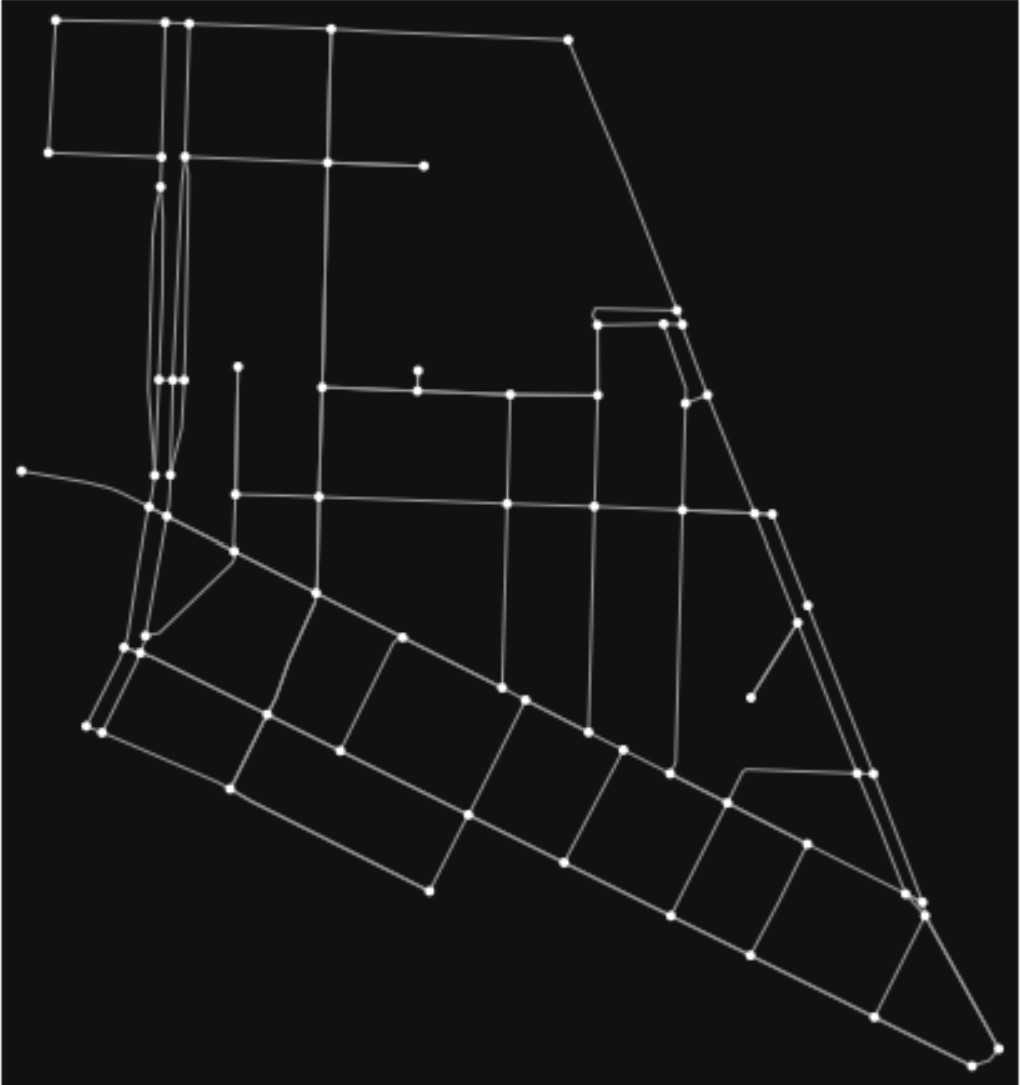}
    \caption{Downtown Brooklyn network (149 nodes, 281 edges).}
    \label{fig:gate1}
\end{figure}

\begin{figure}[h!]
    \centering
    \includegraphics[width=6cm]{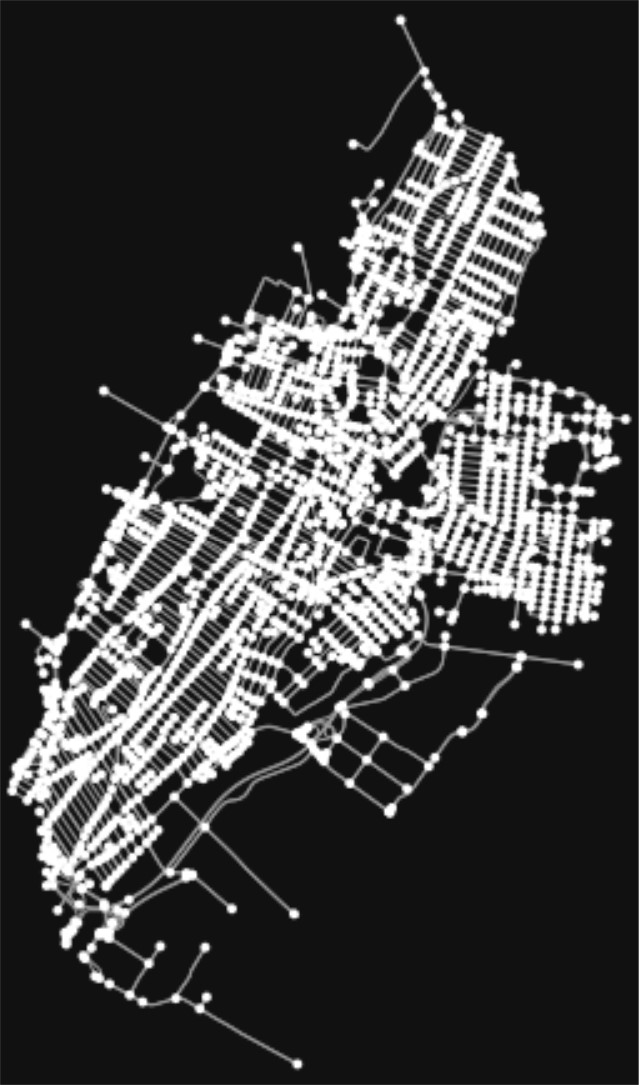}
    \caption{Jersey City network (2232 nodes, 5320 edges).}
    \label{fig:gate2}
\end{figure}

\begin{figure}[h!]
    \centering
    \includegraphics[width=7cm]{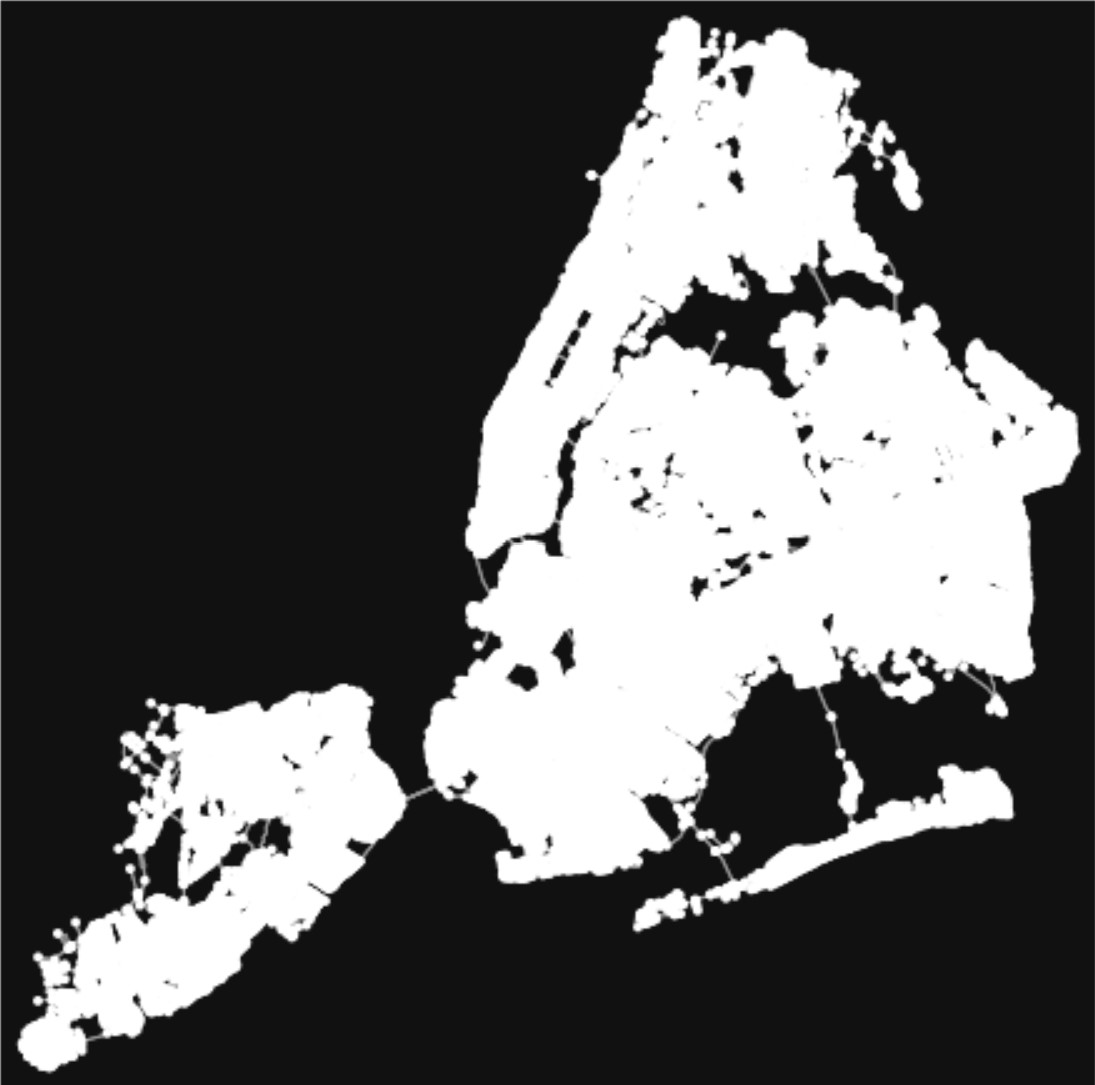}
    \caption{New York City network (55136 nodes, 141085 edges).}
    \label{fig:gate3}
\end{figure}

\newpage
~\newpage
\section{Nodes Expanded vs. $k$ Graphs}

\begin{figure}[h!]
    \centering
    \includegraphics[width=7cm]{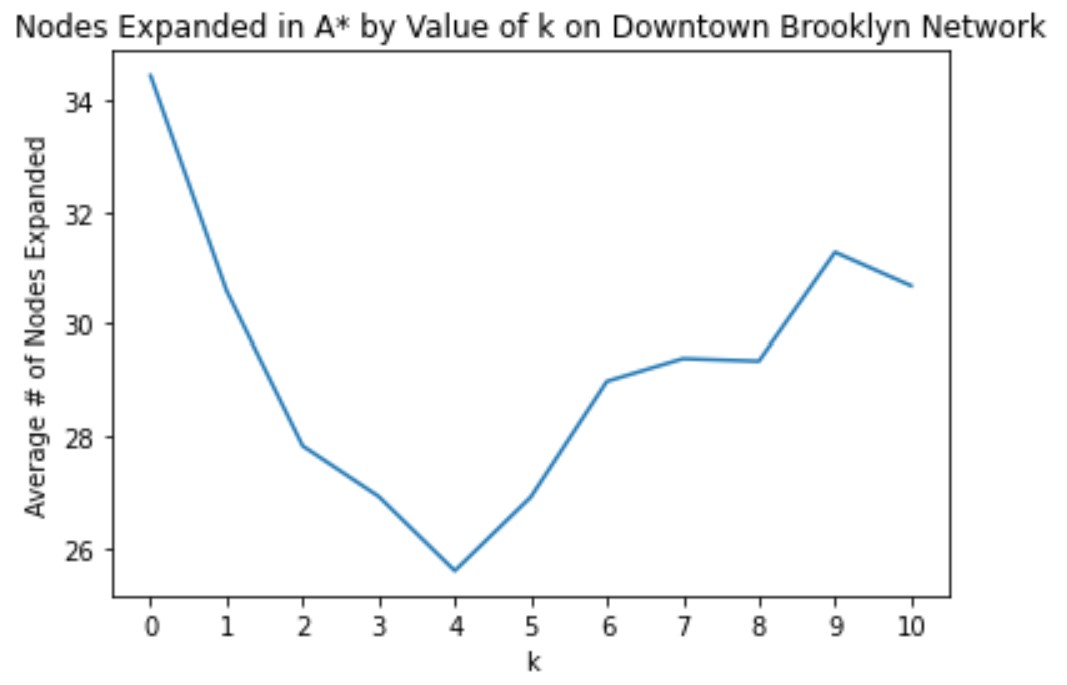}
    \caption{Average number of nodes expanded as a function of $k$ on the Downtown Brooklyn network. Minimal nodes are expanded at $k=4$.}
    \label{fig:gate4}
\end{figure}

\begin{figure}[h!]
    \centering
    \includegraphics[width=7cm]{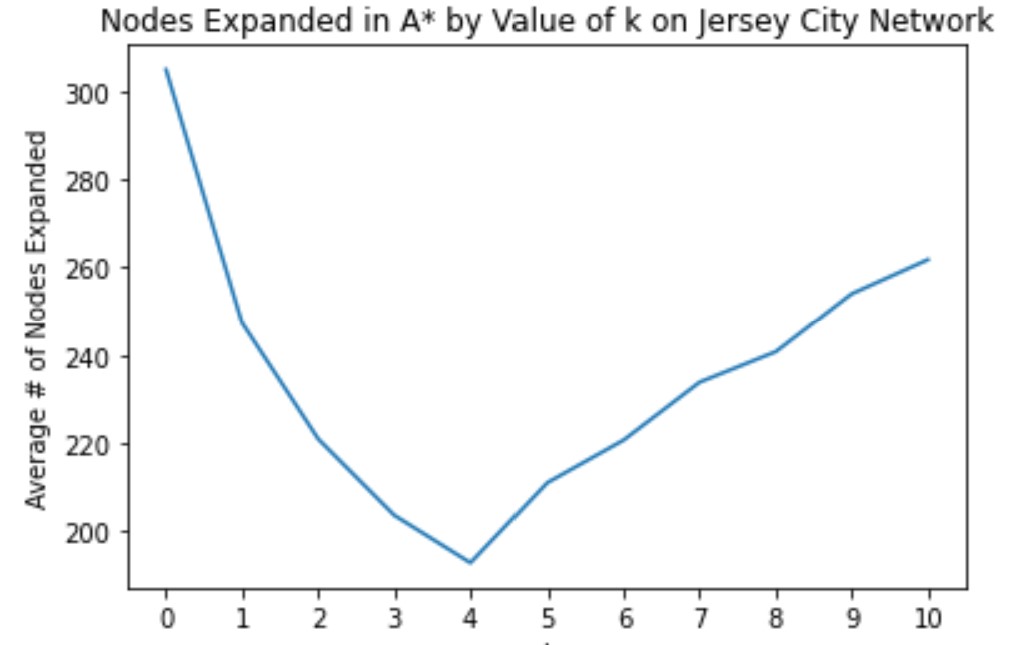}
    \caption{Average number of nodes expanded as a function of $k$ on the Jersey City network. Minimal nodes are expanded at $k=4$.}
    \label{fig:gate6}
\end{figure}

\begin{figure}[h!]
    \centering
    \includegraphics[width=7cm]{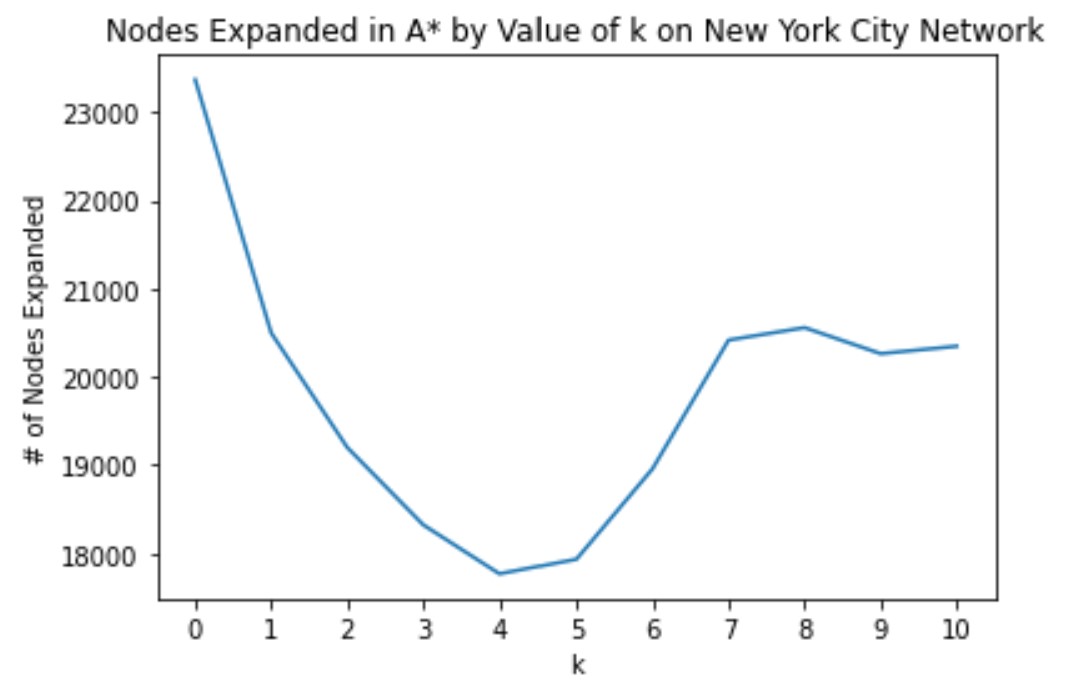}
    \caption{Average number of nodes expanded as a function of $k$ on the New York City network. Minimal nodes are expanded at $k=4$.}
    \label{fig:gate5}
\end{figure}

\end{document}